\documentstyle[emulateapj,amsfonts,psfig,natbib,amsmath]{article}
\def\grb{GRB\,050904}

\def\nrao{1}
\def\cit{2}
\def\uh{3}
\def\bridge{4}
\def\ociw{5}
\def\prince{6}
\def\hubble{7}
\def\anu{7}
\def\psu{8}
\def\srl{9}
\def\gem{10}

\begin{document}

\title{\large An Energetic Afterglow From A Distant Stellar Explosion}

\author{D.~A.~Frail\altaffilmark{\nrao},
P. B. Cameron\altaffilmark{\cit},
M.~Kasliwal\altaffilmark{\cit},
E. Nakar\altaffilmark{\bridge},
P.~A.~Price\altaffilmark{\uh},
E.~Berger\altaffilmark{\ociw,}\altaffilmark{\prince,}\altaffilmark{\hubble},
A.~Gal-Yam\altaffilmark{\cit,}\altaffilmark{\hubble},
S.~R.~Kulkarni\altaffilmark{\cit},
D.~B.~Fox\altaffilmark{\psu},
A.~M.~Soderberg\altaffilmark{\cit},
B.~P.~Schmidt\altaffilmark{\anu},
E.~Ofek\altaffilmark{\cit},
and S.~B.~Cenko\altaffilmark{\srl}
}

\altaffiltext{\nrao}{National Radio Astronomy Observatory, Socorro,
NM 87801}

\altaffiltext{\cit}{Division of Physics, Mathematics and Astronomy,
105-24, California Institute of Technology, Pasadena, CA 91125}

\altaffiltext{\bridge}{Theoretical Astrophysics, California Institute
  of Technology, MS 130-33, Pasadena, CA 91125}

\altaffiltext{\ociw}{Observatories of the Carnegie Institution
of Washington, 813 Santa Barbara Street, Pasadena, CA 91101}
 
\altaffiltext{\prince}{Princeton University Observatory,
Peyton Hall, Ivy Lane, Princeton, NJ 08544}
 
\altaffiltext{\hubble}{Hubble Fellow}

\altaffiltext{\uh}{Institute for Astronomy, University of Hawaii, 
2680 Woodlawn Drive, Honolulu, HI 96822}

\altaffiltext{\anu}{Research School of Astronomy and Astrophysics, 
Australian National University, Mt Stromlo Observatory, via Cotter Rd,
Weston Creek, ACT 2611, Australia}

\altaffiltext{\psu}{Department of Astronomy and Astrophysics, 
Pennsylvania State University, 525 Davey Laboratory, University 
Park, PA 16802}

\altaffiltext{\srl}{Space Radiation Laboratory, MS 220-47, California 
Institute of Technology, Pasadena, CA 91125}

\altaffiltext{\gem}{Gemini Observatory, 670 N. Aohoku Place Hilo, HI 
96720}

\begin{abstract}
  We present the discovery of radio afterglow emission from the high
  redshift ($z=6.295$) burst \grb. The peak flux density for this
  burst is similar to typical low-redshift gamma-ray bursts (GRB). We
  further show that beyond a redshift of order unity, the flux density
  of radio afterglows are largely insensitive to redshift, consistent
  with predictions.  By combining the existing X-ray, near-infrared
  and radio measurements, we derive estimates for the kinetic energy
  and opening angle of the blast wave, and for the density of the
  circumburst medium into which it expands. Both the kinetic and
  radiated energy indicate that \grb\ was an unusally energetic burst
  (10$^{52}$ erg). More importantly, we are able to make an {\it in
    situ} measurement of the density structure of the circumburst
  medium. We conclude that \grb\ exploded into a constant density
  medium with $n_\circ$=680 cm$^{-3}$, which is two orders of
  magnitude above the nominal value for low-redshift GRBs. The next
  generation of centimeter (EVLA) and millimeter radio instuments
  (ALMA) will be able to routinely detect events like \grb\ and use
  them to study magnetic fields, and the atomic and molecular gas in
  the high redshift Universe.

\end{abstract}


\section{Introduction}\label{sec:intro}

Understanding the reionization of the Universe, when the first
luminous sources were formed, is one of the lastest frontiers of
observational cosmology. Constraints have been obtained using
diagnostics such as quasar studies of the Gunn-Peterson absorption
trough, the luminosity evolution of Ly$\alpha$ galaxies, and the
polarization isotropy of the cosmic microwave background. Taken
together, these data portray a complicated picture in which
reionization has taken place over a wide range of redshifts (6$\leq
z\leq 20$) rather than at one specific epoch.  The dominant source of
reionization appears to be due to ultraviolet emission from young,
massive stars (see review by \citealt{fck06}).

As the most luminous explosions in the Universe, gamma-ray bursts
(GRBs) are potential signposts of these early massive stars. The
radio, infrared and X-ray afterglow emission from GRBs are in
principle observable out to z$\sim 30$
\citep{mir98,lr00,cl00b,gmaz04,im05}. It is estimated that 10\% of
GRBs detected by the {\it Swift} satellite are at $z>5$
\citep{nah+05,bl06}.  The most distant GRB to date is \grb\ at
$z=6.295$. It was detected by the {\it Swift} Burst Alert Telescope
and localized to an accuracy of a few arcseconds by the {\it Swift}
X-ray Telescope \citep{gcg+04}.  Follow-up optical and near-infrared
observations (NIR) were begun shortly thereafter.  Details on the
discovery and properties of the X-ray and NIR afterglow of this burst
can be found in several papers \citep{cus06,hnr+06,bad+06}. An
optical/NIR spectrum taken by the {\it Subaru} telescope 3.4 days
after the burst showed multiple heavy metal absorption lines and a
Gunn-Peterson trough with a damping wing redward of the Ly$\alpha$
cutoff which was used to derive the neutral fraction $x_H<0.6$
\citep{tkk+06,kka+06}. With a {\it Swift} GRB detection rate of 100
yr$^{-1}$, GRBs could one day replace quasars as the preferred probe
of the high redshift Universe.

In this paper we report on the detection of the radio afterglow from
\grb\ (\S\ref{sec:obs}), which makes it possible to derive physical
properties of the explosion and the circumburst medium
(\S\ref{sec:afterglow}). Our results are compared with predictions for
the properties of the explosion and how the progenitors of high
redshift GRBs are expected to have shaped their surrounding
environments (\S\ref{sec:disc}).

\section{Observations and Results}\label{sec:obs}

Radio observations were undertaken with the Very Large
Array\footnote{The Very Large Array is operated by the National Radio
  Astronomy Observatory, a facility of the National Science Foundation
  operated under cooperative agreement by Associated Universities,
  Inc.} (VLA) at a frequency of 8.46 GHz (see Table~\ref{tab:vla}). To
maximize sensitivity, the full VLA continuum bandwidth (100 MHz) was
recorded in two 50 MHz bands.  Data reduction was carried out
following standard practice in the {\it AIPS} software package. Some
additional phase and amplitude self-calibration was necessary to
remove the contaminating effects of a bright (10 mJy) radio source
four arcminutes northeast from the GRB.

Radio emission from \grb\ was not detected during the first week,
ruling out a bright short-lived component similar to GRB\,990123
\citep{kfs+99}. By averaging the data from this period, we obtain a
peak flux at the position of the NIR afterglow of 33$\pm$14 uJy. A
clear detection was made during three epochs 34-37 days after the
burst (see Table~\ref{tab:vla}).  Averaging the data taken in October
we obtain a mean flux density of 76$\pm$14 uJy. The spectral radio
luminosity, expressed as $L_\nu=4\pi
F_\nu\,d_L^2\,(1+z)^{\alpha-\beta-1}$ = 2.5$\times 10^{31}$ erg
s$^{-1}$ Hz$^{-1}$ (where $F_\nu\propto{t^\alpha}\nu^{\beta}$ and
$\alpha\sim 0$ and $\beta=1/3$ has been assumed, corresponding to an
optically thin, flat post-jet break light curve), is normal for GRBs
\citep{skb+04}, and is two orders of magnitude below the highest
redshift radio-loud quasars \citep{cwb+04}.

\begin{figure}
\centerline{\psfig{file=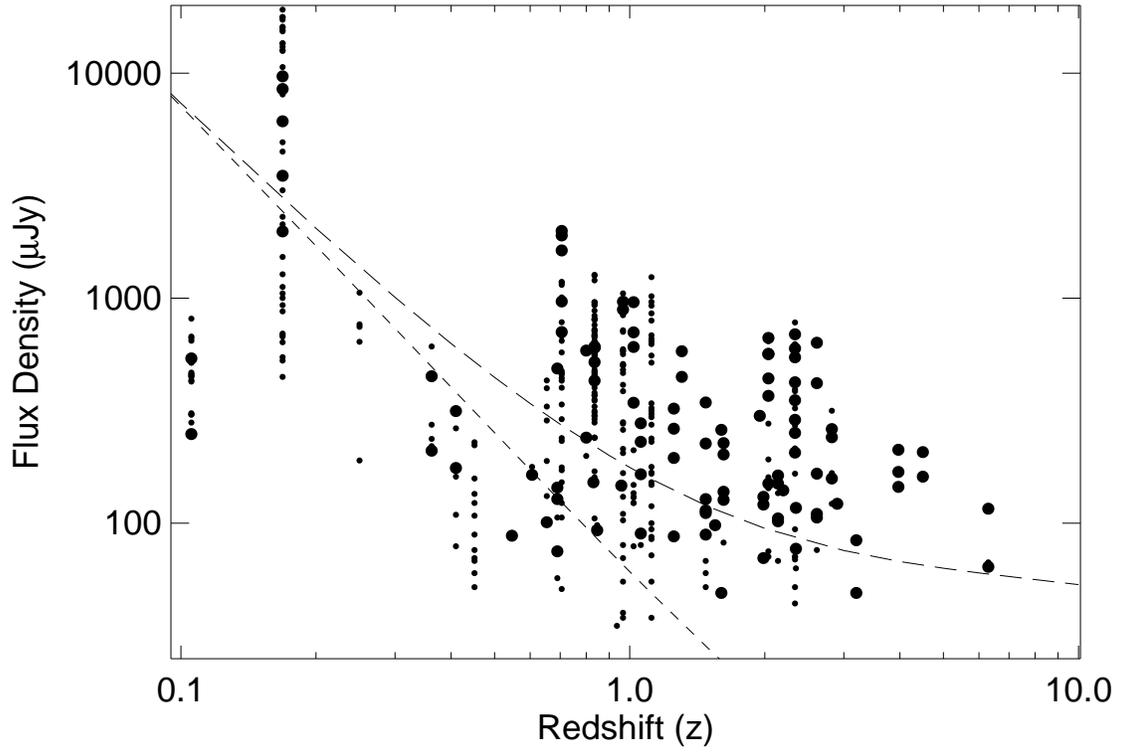,width=6in}}
\caption[]{Radio flux density versus redshift for a complete
  sample of 60 bursts observed at radio wavelengths from 1997 to 2006.
  The radio measurements were made at 8.5 GHz and are taken from
  \cite{fkbw03} and the public radio afterglow database
  ({http://www.aoc.nrao.edu/$\sim$dfrail/grb\_public.shtml}).  For
    clarity only the 42 GRBs with known redshifts and radio afterglow
    detections are plotted. The large (small) circles indicate
    measurements taken less (more) than 5 days before (after) the
    burst in its rest frame. The long dashed line is the flux density
    for a canonical afterglow model at $\Delta t$=5 days (in the
    observer frame). It assumes $E_{k,iso,52}=10^{53}$ erg,
    $\theta_j=0.1$ rad, $n_\circ=10$ cm$^{-3}$, $p=2.2$,
    $\epsilon_e$=0.1, and $\epsilon_B$=1\% (see text for more
    details).  The short dashed line shows the expected decrease if
    the flux scales simply as the inverse square of the luminosity
    distance.
\label{fig:data}}
\end{figure}

We show these detections in Fig.~\ref{fig:data} together with a
complete sample of 8.5 GHz flux density measurements for GRBs with
known redshifts. Neither the time-to-peak nor the flux density of
\grb\ is unusual compared to the known sample of GRBs at lower
redshifts. In the source rest frame the flux density for the majority
of afterglows is reached before 5 days postburst. More interestingly,
the average centimeter flux density in Fig.~\ref{fig:data} shows only
a weak dependence on redshift. 

This effect was predicted by \cite{cl00b} and is shown here for the
first time.  It is reminiscent of the ``negative k-correction'' for
submillimeter observations of ultraluminous infrared galaxies
\citep{bsi+02}. The afterglow flux density remains high because of the
dual effects of spectral and temporal redshift, offsetting the dimming
due to the increase in distance \citep{lr00}. We illustrate in
Fig.~\ref{fig:data} how slowly the centimeter flux density decreases
for a canonical GRB afterglow beyond $z\sim 1$ due to this effect
(long dashed lines), compared to one whose luminosity is assumed
constant (short dashed lines). Observational bias does not explain the
flattening in Fig.~\ref{fig:data} because the detection rate of radio
afterglows is largely insensitive to redshift. Of the 60 GRBs with
known redshift that have been observed in the radio, there are 42 with
detected afterglows. Their detection rate above and below $z=1$ is
identical.

%{sec:obs}The subsequent decline is
%likely the result of a short-lived reverse shock or the passage of the
%synchrotron peak frequency (or self-absorption frequency) through the
%band. The

\section{Afterglow Modeling}\label{sec:afterglow}

%{\it quantify the density in the other model and the energy range in the
%best fit model.}

We now proceed to combine these radio data (\S{\ref{sec:obs}}) with
the extensive X-ray and optical/NIR data for \grb\ in order to
constrain the physical parameters of the outflow and the circumburst
medium. We interpret these multi-wavelength data within the framework
of the relativistic blastwave model (see \citealt{mes02} for a
review).  The particular fitting approach that we take is described in
more detail in \cite{yhsf03}.

To limit the number of free parameters in the fit, we chose to model
only the evolution of the forward shock after energy injection has
ceased. The early X-ray light curve shows complex flaring activity,
common in {\it Swift} bursts but lasting an unusually long time
($\Delta t<0.7$ days), corresponding to a central engine lifetime of
more than 2 hrs in the source rest frame \citep{cus06}. On-going
energy injection is also suspected based on the flares and the
flattening of the optical/NIR lightcurves on similar timescales
\citep{bad+06,hnr+06}. If these early data are excluded the NIR data
show a smooth power-law evolution with a clear break at $t_j=2.6 \pm
1.0$ days and decay indices before and after the break of
$\alpha_1=-0.72{^{+0.15}_{-0.20}}$ and $\alpha_2=-2.4\pm{0.4}$
\citep{tac+05}. If interpreted as a jet break, the sharpness of the
transition makes it unlikely that this burst occurred in a wind-blown
environment \citep{kp00} and motivates our modeling choice of a
constant density circumburst medium.

% must put parms in figure caption

\begin{figure}
\centerline{\psfig{file=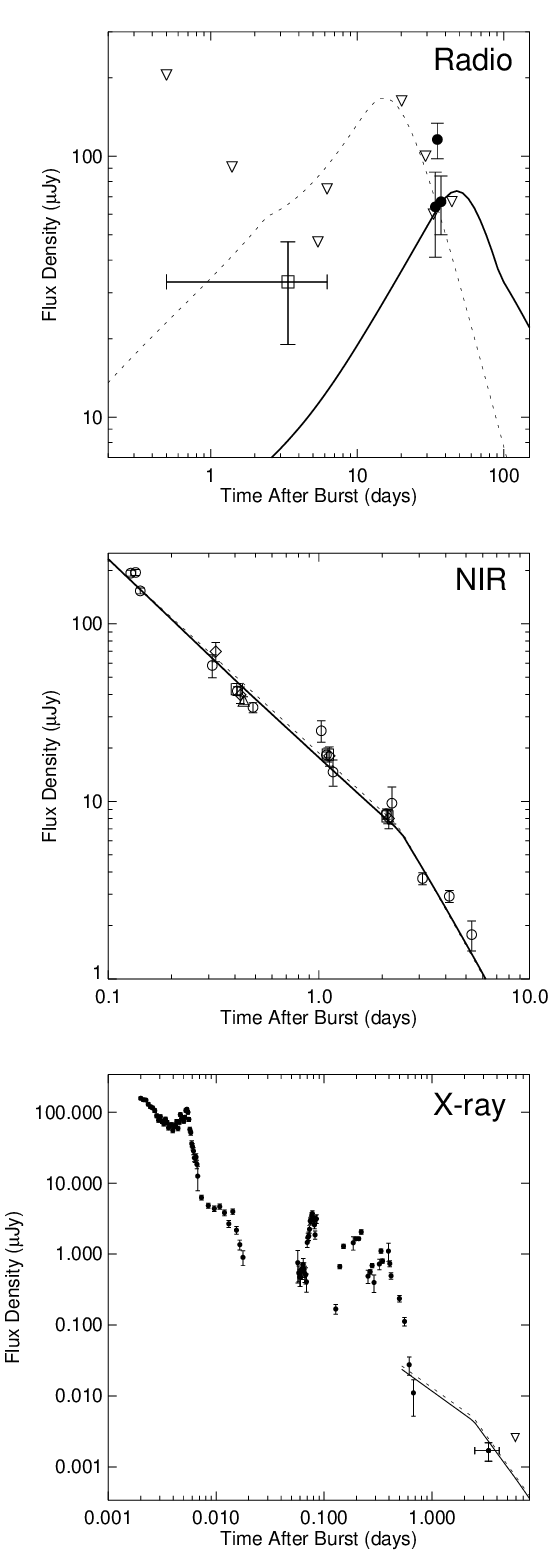,width=3.0in}}
\vspace{-0.4cm}
\caption[]{Broad-band afterglow fits for the radio, near infrared 
  (NIR) and X-ray data of \grb. The radio data are taken from
  \S\ref{sec:obs}. Solid circles indicate detections, upper limits are
  inverted triangles (plotted as flux + 2$\sigma$). The open square is
  the peak flux obtained by averaging the first week of data. The NIR
  data are taken from \cite{tac+05} and \cite{hnr+06}, with a small
  correction made for Galactic dust extinction \citep{sfd98} before
  converting to flux density \citep{bcp98}. JHK and K$^\prime$ are
  shown for display purposes on the same plot after scaling by the
  best-fit spectral index. We have converted the {\it Swift} X-ray
  data ( 0.2-10 keV) to flux density using the average photon index
  $\Gamma=-1.84$ and a frequency of 2.8$\times 10^{17}$ Hz (see
  \citealt{cus06}). The solid line is our best-fit forward shock model
  for a constant density circumburst medium with
  $E_{k,iso,52}$=8.8$\times 10^{53}$ erg, $\theta_j$=0.14 rad,
  $n_\circ$=680 cm$^{-3}$, $p$=2.14, $\epsilon_e$=2.0\%, and
  $\epsilon_B$=1.5\% (see text for more details).  The thin dashed
  line shows the effect of model fitting when the value of $n_\circ$
  is fixed at 100 times lower than the best-fit value.
\label{fig:model}}
\end{figure}

Our best-fit forward shock model is shown in Fig.~\ref{fig:model}.  We
fit for the isotropic kinetic energy of the shock $E_{k,iso,52}$, the
opening angle of the jet $\theta_j$, the density of the circumburst
medium $n_\circ$, the electron energy index $p$, and the fraction of
the shock energy density in relativistic electrons $\epsilon_e$ and
magnetic fields $\epsilon_B$.

Our model makes the unusual prediction that the synchrotron cooling
frequency lies below {\it both} the X-ray and optical bands before the
first J-band detection of the afterglow at $\Delta t$=3.07 hrs
\citep{hnr+06}. This implies a steep spectral slope of
$\beta=-p/2=-1.07$ \citep{sph99}.  This is likely the case at $\Delta
t$=3 days where we derive an NIR/X-ray slope of $\beta_{ox}=-1.1\pm
0.05$. This value also agrees with \cite{tac+05} who derive a best fit
optical/NIR spectral index of $\beta_o=-1.25\pm 0.25$ over the
interval from 0.4 to 7 days postburst.  A steep spectral slope of
$\beta_x\sim -0.85$ is seen even when the X-ray light curve is highly
time variable between 0.05$\leq \Delta t \leq 0.7$ days.  Thus we find
that the available data support a small value for the cooling
frequency.

Our best-fit model favors a circumburst density $n_\circ$=680
cm$^{-3}$. To test the robustness of this result we ran several model
fits, fixing $n_\circ$ over a range of values between 0.7 cm$^{-3}$
and 7$\times 10^4$ cm$^{-3}$. As expected, the X-ray and NIR data are
relatively insensitive to $n_\circ$, so the radio afterglow provides
the tightest constraints. Large densities $n_\circ> 10^{4}$ cm$^{-3}$
are ruled out since the solutions are bounded by the condition that
the magnetic and electron densities do not exceed equipartition.  Low
densities $n_\circ< 10$ cm$^{-3}$ are also not favored since they
predict bright early radio emission in the first week well in excess
of that which is observed (\S\ref{sec:obs}). Over the allowable
density range, $E_{k,iso,52}$ varies from 40 to 530, and therefore it is
no better constrained by the afterglow data than the {\it isotropic}
radiated energy $E_{\gamma,iso,52}$ is by the prompt emission
\citep{cus06}.

%Such
%large densities may also imply substantial extinction, which is not
%supported by the optical/NIR data \citep{tkk+06,tac+05}. 

Another possibility is that radio afterglow flux density measurements
are biased by the turbulent ionized interstellar medium of our Galaxy
which induces non-Gaussian intensity fluctuations for compact radio
sources.  Indeed, it has been shown by \cite{cl91} that repeated short
observations similar to those in Table~\ref{tab:vla} increase the odds
of detecting a weak radio signal close to the detection threshold. We
replaced the detections with upper limits and re-ran the fit. High
density solutions are still preferred ($n_\circ\simeq 10^{4}$
cm$^{-3}$) because both the X-ray and optical afterglows of \grb\ were
bright \citep{cus06,hnr+06}.  Thus, in order to suppress the radio
emission at 8.5 GHz (or 62 GHz in the rest frame), a large value of
the synchrotron self-absorption is required and hence a large
circumburst density.

%Alternatively, since we lack
%concurrent broadband data at $\Delta t \sim 35$ days, the radio
%emission may not originate from the forward shock

\section{Discussion}\label{sec:disc}

With the detection of the radio afterglow from \grb\ we have a
complete multi-wavelength dataset for this high redshift burst.  By
comparing the physical properties of the burst and its circumburst
environment to those GRBs found at lower redshifts, we hope to gain
some insight into the birth and death of the earliest generations of
stars in the Universe.

%The afterglow of \grb\ was not exceptionally luminous. Its radio
%spectral luminosity, expressed as $L_\nu=4\pi
%F_\nu\,d_L^2\,(1+z)^{\alpha-\beta-1}$ = 2.5$\times 10^{31}$ erg
%s$^{-1}$ Hz$^{-1}$ (where $\alpha\sim 0$ and $\beta=1/3$ has been
%assumed, corresponding to an optically thin, flat post-jet break light
%curve), is normal for GRBs \citep{skb+04}, and is two orders of
%magnitude below the highest redshift radio-loud quasars
%\citep{cwb+04}. Likewise, neither the X-ray flux nor the X-ray
%luminosity are unusual compared to the sample of GRBs at $z=1-2$
%\citep{bkf03,cus06}. The optical afterglow was not exceptionally
%luminous except for a short-lived optical flare discovered by
%\cite{bad+06} that coincided with an X-ray flare at $\Delta t\simeq
%500$ s. Its luminosity was similar to another optical flare seen
%towards GRB\,990123 \citep{abb+99} at $z=1.6$, and is likely the
%result of a rare occurence of a bright, short-lived reverse shock
%\citep{sp99b}.

There is good support for interpretating \grb\ as an energetic event.
From our afterglow modeling (\S\ref{sec:afterglow}) of the isotropic
kinetic energy of the shock $E_{k,iso,52}=88$ (normalized in units of
10$^{52}$ erg). Estimates obtained for the {\it isotropic} radiated
energy \citep{cus06} with $66 \leq E_{\gamma,iso,52} \leq 320$.  The
quoted range in $E_{\gamma,iso,52}$ relects the uncertainty in the
location of the peak of the gamma-ray spectrum. These constraints can
be refined owing to our fitting the jet opening angle $\theta_j$=0.14
rad (see also \citealt{tac+05}). The geometrically corrected energies
are $E_{k,52} \simeq E_{k,iso,52}\times \theta_j^2/2$=0.9, and $0.7
\leq E_{\gamma,52} \leq 3$. Both the radiated and kinetic energy of
this event lie at or beyond the values derived for a large sample of
lower redshift afterglows \citep{pk01,pk02,yhsf03,fb05}. An
independent check on this result uses the X-ray light curve to compute
the geometrically-corrected X-ray luminosity at some fiducial time
(usually taken at $\Delta t$=10 hrs). We derive L$_x$=2.6$\times
10^{45}$ erg s$^{-1}$, again larger than any previous GRB afterglow
\citep{bkf03}.

Perhaps the most interesting aspect of these data is that they allow
us to make an {\it in situ} measurement of the density structure for a
massive star in the early Universe. From our afterglow modeling we
conclude that \grb\ exploded into a constant density medium with
$n_\circ$=680 cm$^{-3}$ -- two orders of magnitude above the nominal
value for lower redshift GRBs \citep{fb05,snbk06}. Line-of-sight
measures \citep{cus06,hnr+06,tkk+06} also indicate that a substantial
column of gas exists towards \grb. The fine-structure silicon lines
SiII$^*$ are the most important since their column density ratios
yield estimates of the electron density $n_e\sim 10^{2.3\pm 0.7}$
cm$^{-3}$ \citep{kka+06}. The close agreement of this line-of-sight
estimate with our {\it in situ} value argues for a common physical
origin for the gas in the immediate vicinity of the GRB.

There are testable predictions for star formation and collapse in the
early Universe (\citealt{bl06} and reference therein).  The first
stars are expected to be very massive and could produce energetic GRB
explosions like \grb.  Likewise, the high density we derived for \grb\ 
may be part of the general increase expected for the ambient density
of the interstellar medium \citep{cl00b}, but more likely just
indicates that this GRB exploded in a dense molecular cloud. A lower
density would be expected if the progenitor of \grb\ was a low
metalicity Wolf-Rayet star \citep{vk05}. However, it remains a
problem, at both low and high redshifts, why the radial density
signature from the collapsar wind is rarely detected \citep{clf04}. A
larger sample is needed before drawing too much from these
predictions.  Deep radio observations will continue to play a role
since can be seen out to the highest redshifts and are sensitive
probes of the total calorimetry of the explosion and the density
structure of the circumburst medium.  Looking further ahead to the
next generation of centimeter (EVLA) and millimeter radio instuments
(ALMA), events like \grb\ will be bright enough to sample primordial
magnetic fields \citep{gbf04}, and probe the cold atomic and molecular
gas from the time of the first light \citep{ioc06}.

%High Density claims...
%Previous claims of high density have been
%made \citep{pgg+01,zka+01}, but they were not subsequently verified by
%multi-wavelength data \citep{yhsf03}.
%mention central engine activiy duration
%Points to RM in he future.
%{\it in situ}

\begin{acknowledgements}

  DAF wishes to thank B. Clark for his generous allocation of VLA time
  during the dynamic scheduling tests.

\end{acknowledgements}

%\bibliographystyle{apj1b}
%\bibliography{journals_apj,/users/dfrail/papers/g021004/radio/mastergrb}
%\bibliography{journals_apj,mastergrb}

\clearpage

\begin{deluxetable}{lrr}
\tablecaption{Radio Observations}
\tablewidth{0pt} \tablehead{
\colhead{Date Obs} & 
\colhead{$\Delta t$} & 
\colhead{$F_{\nu}$\tablenotemark{a}} 
\\ 
\colhead{(UT)} & 
\colhead{(days)} & 
\colhead{($\mu$Jy)} }
\startdata
2005 Sep.~4.58  &  0.50 & 89$\pm$58 \\ 
2005 Sep.~5.48  &  1.40 & 41$\pm$25 \\
2005 Sep.~9.48  &  5.40 & $-$3$\pm$25 \\
2005 Sept.~10.31&  6.23 & 27$\pm$24 \\
2005 Sep.~24.20 & 20.12 & 89$\pm$37 \\
2005 Oct.~3.36  & 29.28 & 40$\pm$30 \\
2005 Oct.~7.30  & 33.22 & $-$10$\pm$35\\
2005 Oct.~8.26  & 34.18 & 64$\pm$23 \\
2005 Oct.~9.34  & 35.26 & 116$\pm$18 \\
2005 Oct.~11.43 & 37.35 & 67$\pm$17 \\
2005 Oct.~18.37 & 44.29 & 13$\pm$27 \\
\enddata
\tablenotetext{a}{All errors are given as $1\sigma$ (rms).}  
\label{tab:vla}
\end{deluxetable}
\clearpage

%Figures

\newpage

\newpage

\end{document}